\documentclass[a4paper,11pt]{article}
\pdfoutput=1 
\usepackage{graphicx}
\usepackage{jinstpub} 

\usepackage{cleveref}
\def\elaser{\ensuremath{e{\rm -laser}}\xspace}
\def\glaser{\ensuremath{\gamma{\rm -laser}}\xspace}
\def\um{\ensuremath{\mu {\textrm m}}\xspace}

\newcommand{\geant}{$\textsc{Geant4}$\xspace}
\usepackage{xspace}
\title{\boldmath Detector Challenges of the strong-field QED experiment LUXE at the European XFEL}

\author[a,1]{A. Santra,\note{Corresponding author.}}

\newcommand{\Ecr}{$\mathcal{E}_{\rm cr}$\xspace}
\newcommand{\epair}{${\rm e}^+{\rm e}^-$\xspace}

\affiliation[a]{Weizmann Institute of Science, \\234, Herzl Street, Rehovot, Israel-7610001}

\emailAdd{arka.santra@weizmann.ac.il}

\abstract{The LUXE experiment aims at studying high-field QED in electron-laser and photon-laser interactions, with the 16.5 GeV electron beam of the European XFEL and a laser beam with power of up to 350 TW. The experiment will measure the spectra of electrons, positrons and photons in expected ranges of 10$^{-3}$ to 10$^9$ per 1 Hz bunch crossing, depending on the laser power and focus. These measurements have to be performed in the presence of low-energy high radiation-background. To meet these challenges, for high-rate electron and photon fluxes, the experiment will use Cherenkov radiation detectors, scintillator screens, sapphire sensors as well as lead-glass monitors for backscattering off the beam-dump. A fourlayer silicon-pixel tracker and a compact electromagnetic tungsten calorimeter with GaAs sensors will be used to measure the positron spectra. The layout of the experiment and the expected performance under the harsh radiation conditions will be presented. Beam tests for the Cherenkov detector and the electromagnetic calorimeter were performed at DESY recently and results will be presented. The experiment received a stage 0 critical approvement (CD0) from the DESY management and is in the process of preparing its technical design report (TDR). It is expected to start running in 2025/6.}

\keywords{LUXE, XFEL}

\arxivnumber{1234.56789} 

\collaboration[c]{on behalf of the LUXE collaboration}

\proceeding{23$^{\text{rd}}$ International Workshop on Radiation Imaging Detectors\\
  26-30 June, 2022\\
  Riva del Garda,Italy}

\begin{document}
\maketitle
\flushbottom

\section{Introduction}
\label{sec:intro}
Quantum Electrodynamics (QED) is well-understood in the perturbative regime. 
However, there is a much less understanding of QED in the strong field regime. 
The strong field regime of QED is characterized by the Schwinger critical field, \Ecr$\equiv m_e^2c^3/e\hbar=1.32\times10^{18}$ V/m with $m_e$ and $e$ are electron mass and charge respectively. 
This is the strength at which an electric field spontaneously produces electron and positron pairs from the vacuum, as originally formulated by Schwinger for static fields.
A static electric field as high as \Ecr is not achievable in the lab, but this regime can be probed in the collision of high intensity laser beams and highly energetic electron or photons. 
Non-linear Compton scattering and laser assisted \epair pair production are two of the strong field phenomena that have been studied extensively theoretically~\cite{Gonoskov:2021hwf,Fedotov:2022ely} which are represented by the following equations respectively:

\begin{subequations}\label{eq:pro}
\begin{align}
\label{eq:pro:1}
e^-+n\gamma_L\rightarrow e^-+\gamma
\\
\label{eq:pro:2}
\gamma+n\gamma_L\rightarrow e^++e^-
\end{align}
\end{subequations}

with $n$ being the number of laser photons ($\gamma_L$) participating in the process.

The interaction between laser electric field and high energy particles can be characterized by two dimensionless parameters, namely dimensionless intensity parameter $\xi$ and quantum non-linearity parameter $\chi$. 
The definition of $\xi$ is as follows:
\begin{equation}
 \xi = \frac{e\mathcal{E}_L}{m_ec\omega_L}=\frac{m_ec^2\mathcal{E}_L}{\hbar\omega_L\mathcal{E}_{\rm cr}}
\end{equation}
where $\mathcal{E}_L$ is the root mean square value of the laser electric field and $\omega_L$ is the frequency of the laser. 
The quantum non-linearity parameter is given by
\begin{equation}
    \chi = \frac{e\hbar}{m_e^3c^5}\sqrt{(F_{\mu\nu} p^\nu)^2}
\end{equation}
with $F_{\mu\nu}$ the field tensor, $p^\nu$ the four momentum of the high energetic particle. 
This can be rewritten with the help of electric field in particle's rest frame ($\mathcal{E}^*$) as
\begin{equation}
    \chi = \frac{\mathcal{E}^*}{\mathcal{E}_{\rm cr}}.
\end{equation}

\begin{figure}[htbp]
\centering 
\includegraphics[width=.4\textwidth,angle=270]{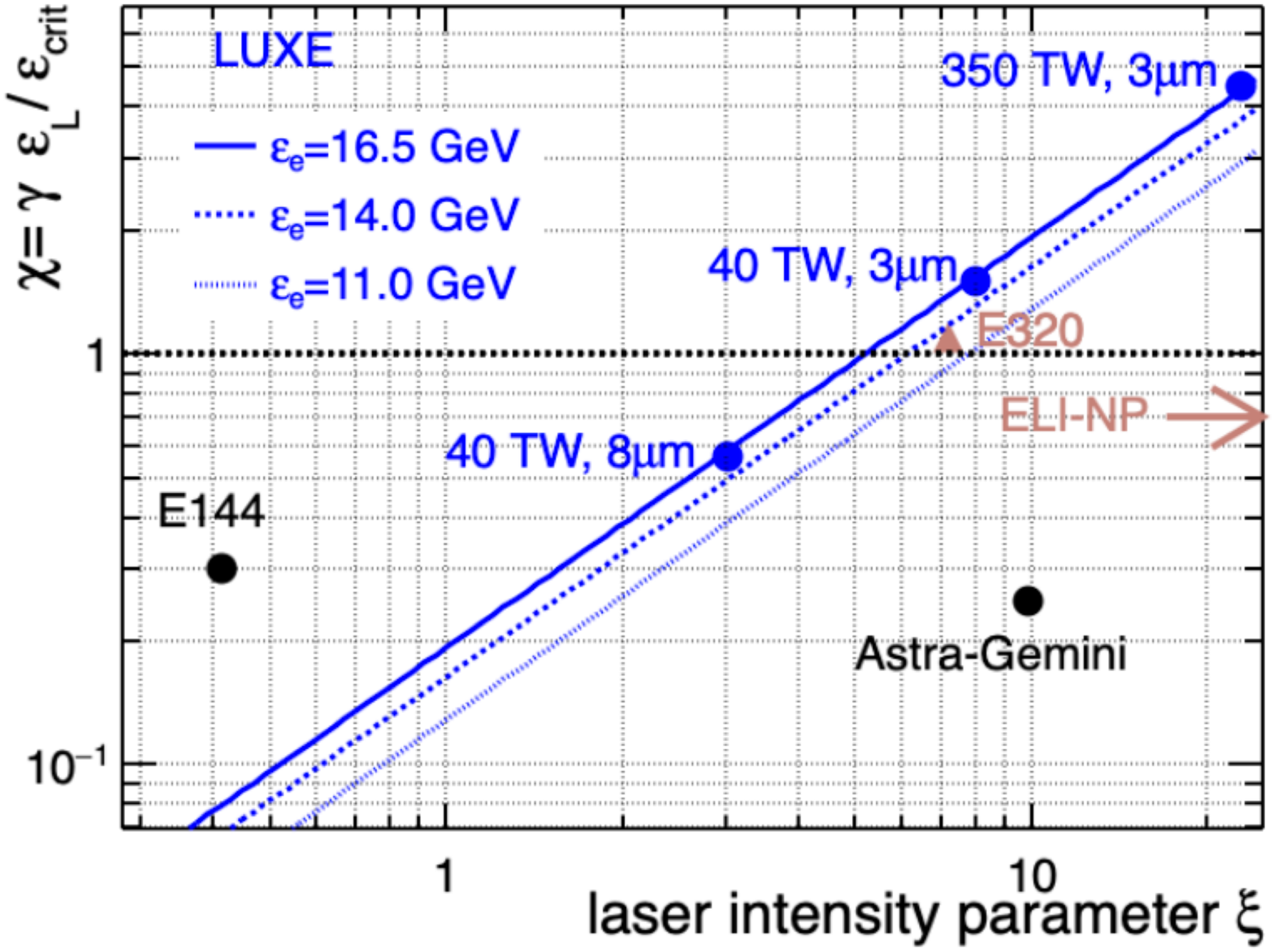}
\caption{The $\chi$ vs $\xi$ parameter space probed by various experiments. The reach of LUXE in the parameter space is shown by three blue lines corresponding to three different electron beam energy. The 40 TW laser focus spot size shows the phase-0 running of LUXE, and 350 TW laser focus spot size shows the phase-1 running of LUXE.}
\label{fig:range}
\end{figure}

The first experiment to study the strong field QED was E144~\cite{Bamber,Reiss:1971wf} based at SLAC.
This experiment studied the non-linear Compton scattering and laser assisted \epair production and could reach up to $\xi=0.34$. 
There are some other experiments in present day that study or propose to study the strong field QED region.
For example,  Astra-Gemini laser facility~\cite{Poder:2017dpw,Cole:2017zca} studied this region.
Future experiments like European Extreme Light Infrastructure (ELI)~\cite{Zamfir:2014msa} and LUXE~\cite{Abramowicz:2019gvx, Abramowicz:2021zja,Hartin:2018sha} are planned. 
The reach of these experiments in the parameter space is summarized in~\cref{fig:range}.
The tremendous progress in the laser technologies in recent days is helping to investigate uncharted territories in $\xi$ and $\chi$ experimentally.

\section{The LUXE experiment}
\label{sec:luxeex}
The final states from~\cref{eq:pro} show that LUXE experiment will detect and measure kinematic properties of positrons, electrons and photons. 
Keeping this in mind, the LUXE setup comprises of three separate detector systems:
\begin{enumerate}
    \item Positron detection system,
    \item Electron detection system,
    \item Photon detection system.
\end{enumerate}
Since the initial particles for~\cref{eq:pro} are different, LUXE will run in both \elaser and \glaser mode.
The focus of this proceeding will be the \elaser setup, but the details about the \glaser setup can be found elsewhere~\cite{Abramowicz:2021zja}. 

\begin{figure}[htbp]
\centering 
\includegraphics[width=.50\textwidth,angle=270]{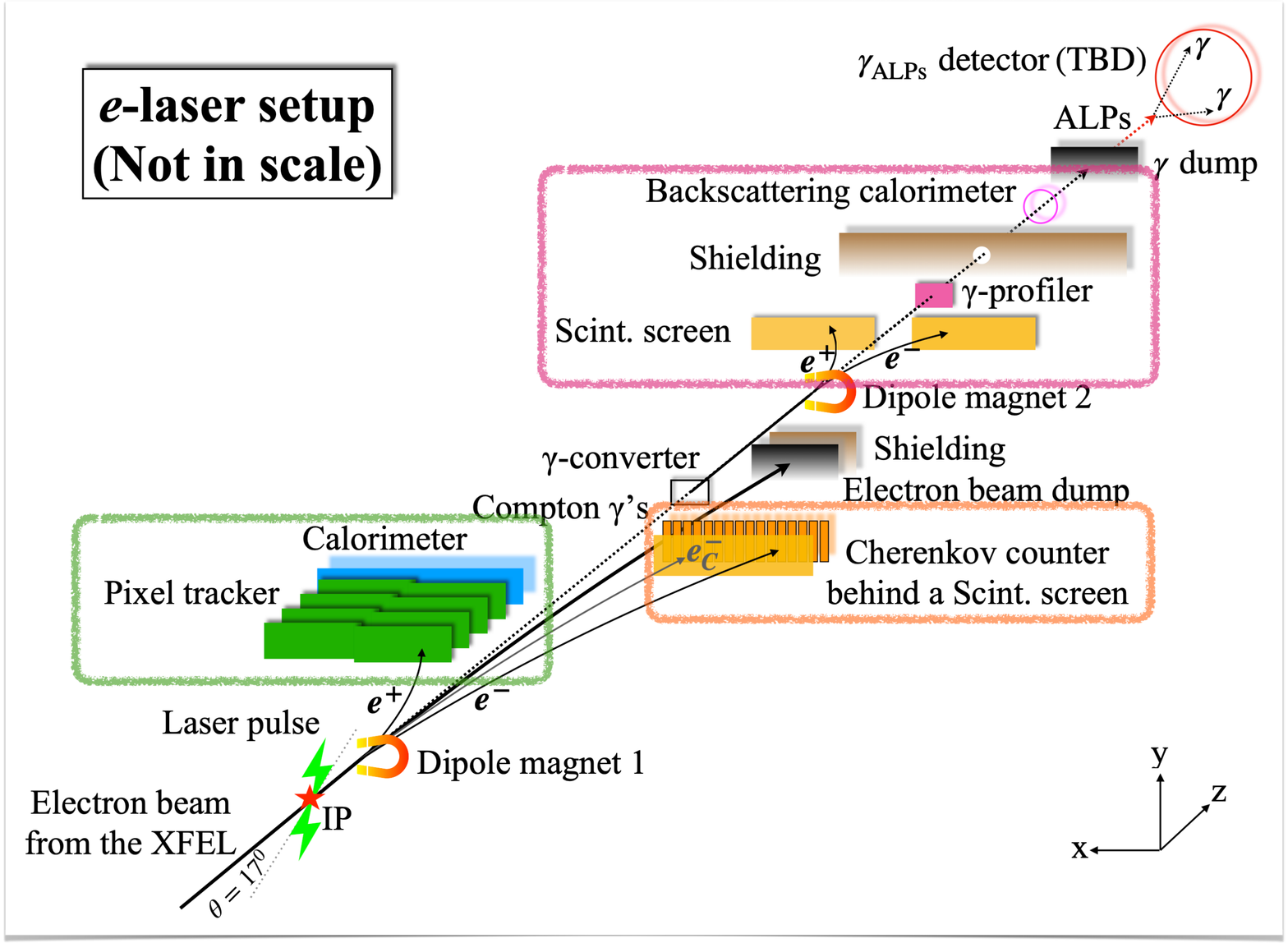}
\caption{LUXE Detector Layout in \elaser mode. The components of the positron detection system iare marked in green, the sub-detectors in electron detection system are highlighted in orange and the photon detection system is shown in pink. The Axion-like particles (ALP) detection system is also shown at the very end, but the technology for this detector system is not finalized yet.}
\label{fig:luxe}
\end{figure}

A schematic diagram of LUXE \elaser setup is shown in~\cref{fig:luxe}. 
In this setup, the high energetic electron beam from XFEL~\cite{Altarelli:2006zza} will collide with the intense laser pulse at the interaction point (IP) located in the interaction chamber. 

\section{The LUXE detector systems}
\label{sec:detsec}

\subsection{Positron detection system}
\label{sec:pds}
The expected number of positrons in LUXE ranges from $10^{-2}$ to $10^5$ per bunch crossing (BX). 
Therefore, the challenge for the positron detection system is to have an excellent positron detection efficiency even in the presence of high background, for a wide range of signal multiplicity.
The background mainly comes from the original electron beam having $1.5\times10^9$ electrons per BX, dumped in the experimental area.
In order to satisfy the needs of LUXE, the positron detection system contains silicon pixel tracker and electromagnetic calorimeter.

\begin{figure}[htbp]
\centering 
\includegraphics[width=.42\textwidth,angle=0]{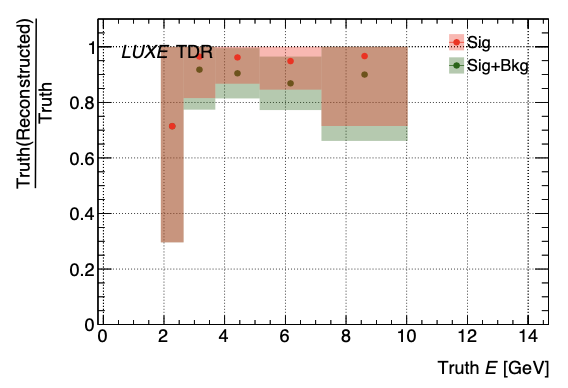}
\caption{The positron reconstruction efficiency as a function of true energy. The effciency here is shown for signal sample with $\xi=3$.}
\label{fig:trackereff}
\end{figure}

The tracker detector will be made of four layers of ALPIDE pixel sensors, developed by ALICE~\cite{ALICE:2018fuj}.
The pixel size is $27\times29$ $\mu$m$^2$ with a spatial resolution of 5~\um. 
The tracking efficiency as a function of true energy for a specific signal sample prepared with \geant~\cite{AGOSTINELLI2003250} is shown in~\cref{fig:trackereff}.
Initial simulation shows that the tracker detector has detection efficiency more than $95\%$ with energy resolution better than $\sim1\%$.

\begin{figure}[htbp]
\centering 
\includegraphics[width=.42\textwidth,angle=0]{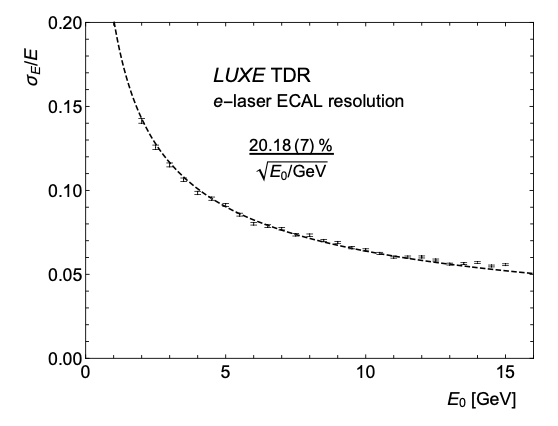}
\caption{The relative energy resolution of the electromagnetic calorimeter as a function of the positron energy.}
\label{fig:ecal}
\end{figure}

The electromagnetic calorimeter of LUXE is envisaged as a sampling calorimeter containing 20 layers of tungsten plate of thickness 3.5 mm (1~$X_0$) and assembled sensor planes placed in between the absorber plates within 1 mm gap.
The whole structure will be supported by an aluminium frame which will have slots on top to accommodate the front-end boards.
The sensors is made of high resistivity GaAs wafers of 500~\um thickness.
Each sensor consists of 150 pads and has a surface area of $7.56\times 5.19$ cm$^2$.
The fiducial volume of the calorimeter will be $53\times5.2\times19$ cm$^3$.
In~\cref{fig:ecal}, the relative energy resolution of the calorimeter is shown.

\begin{figure}[htbp]
\centering 
\includegraphics[width=.48\textwidth,angle=270]{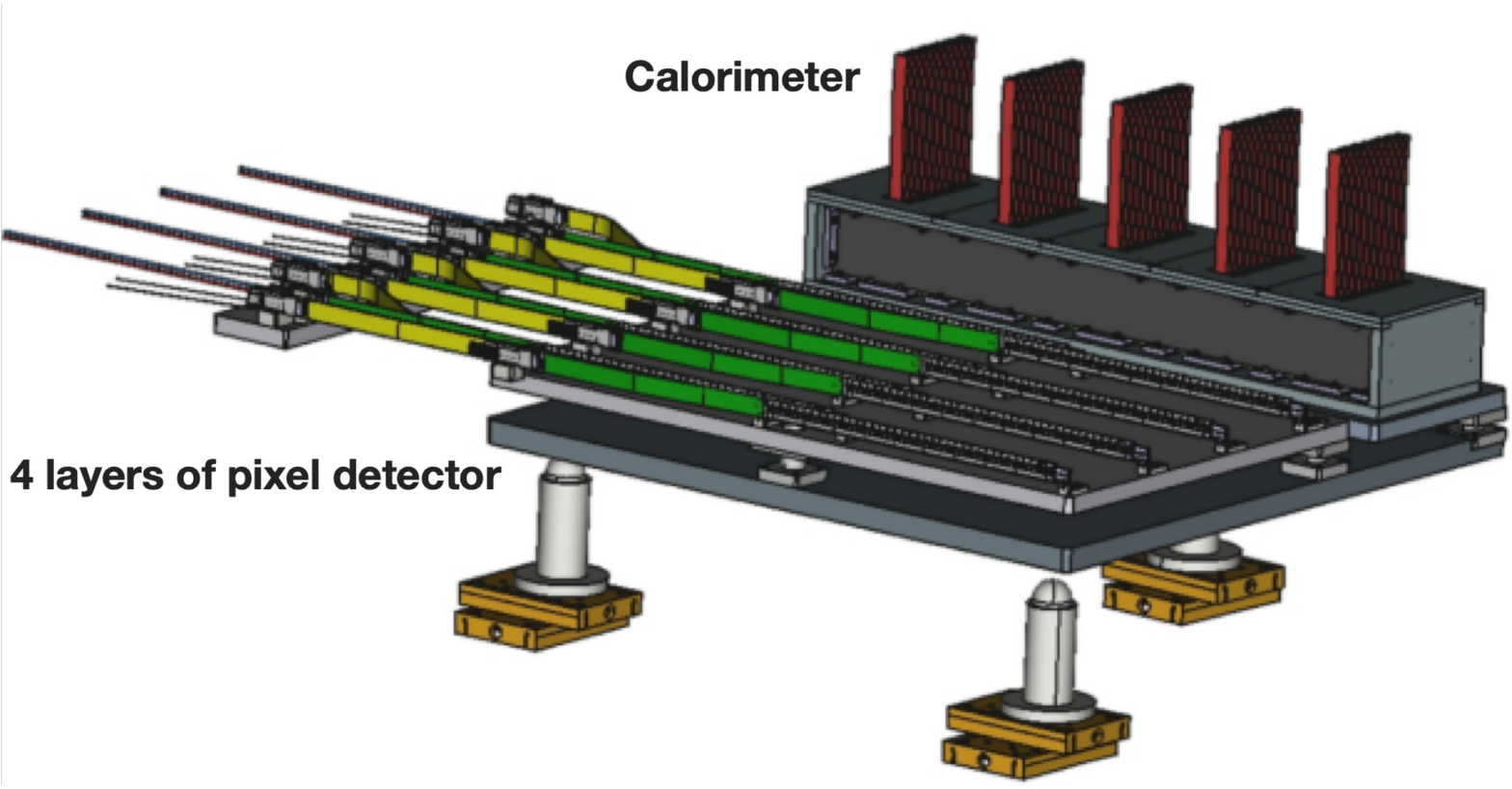}
\caption{The positron detection system of LUXE experiment. Here four layers of tracker are followed by a electromagnetic calorimeter. The front end electronics will be positioned on top of electromagnetic calorimeter.}
\label{fig:pds}
\end{figure}

A sketch of the pixel tracker and electromagnetic calorimeter is shown in~\cref{fig:pds}.

\subsection{Electron detection system}
\label{sec:eds}
\begin{figure}[htbp]
\centering 
\includegraphics[width=.45\textwidth,angle=270]{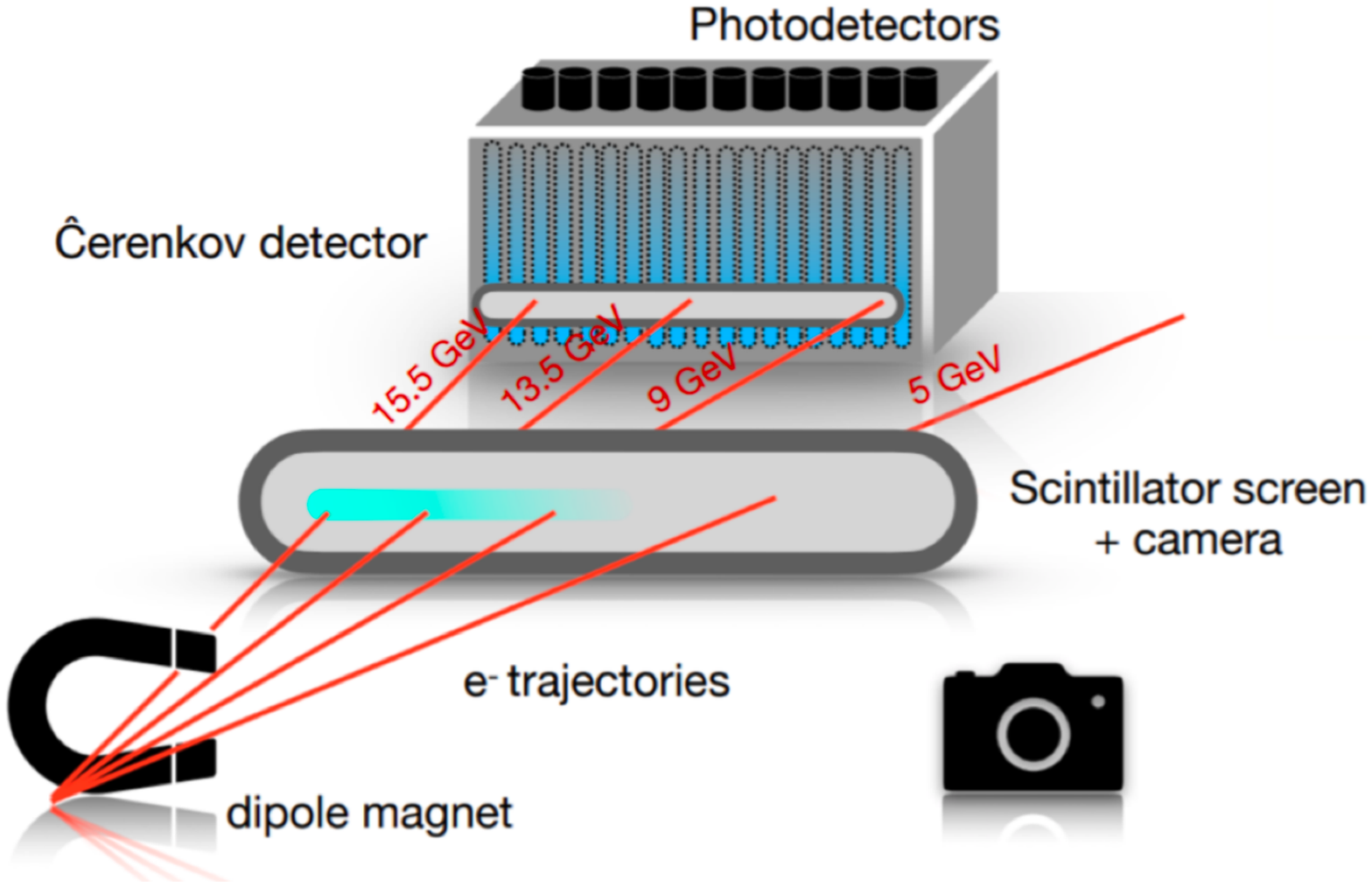}
\caption{The electron detection system of LUXE comprising of scintillator screen and Cherenkov detector. The light profile coming out of the screen is imaged with an optical camera. The light produced in the Cherenkov medium will be reflected to an array of photodetectors.}
\label{fig:ele}
\end{figure}

The electron detection system of LUXE is made of a Cherenkov detector behind a scintillator screen.
The Cherenkov detectors are capable of detection of electrons with high particle flux, of the order of $10^4$-$10^9$ electrons per BX which will be seen in the LUXE \elaser mode.
The scintillator screen is capable of high position resolution and will be used mainly to measure the low particle flux and low energy spectrum of electrons.
The scintillator screen is made of Tb-doped Gadolinium Oxysulfide, which is very radiation hard (up to 10 MGy).
The scintillator screens will be coupled with CMOS camera which will take picture when light emits from the scintillator screen. 
This technology provides position resolution of $\mathcal{O}(100$\um) at $\sim50$ MeV.
On the other hand, the Cherenkov detectors offer a greater resistance to low-energy background particles, and are useful for the challenges of high particle flux if adequate refractive medium is selected.
The Cherenkov detector at LUXE will contain reflective straw tube channels filled with air as an active medium. 
A schematic diagram of the scintillator screen and Cherenkov detector is shown in~\cref{fig:ele}.

\subsection{Photon detection system}
\label{sec:phds}
The photons that are produced at the IP propagate upstream undeflected.
The photon detection system comprising of Gamma Spectrometer, Gamma Profiler and Gamma Flux Monitor will measure the photon energy spectrum, angular distribution and the photon flux respectively. 
The Gamma Spectrometer has a small tungsten target which converts photons into e$^+$e$^-$ pairs and then there is a dipole magnet to separate e$^-$ and e$^+$. 
The energy of e$^-$ and e$^+$ is measured by Tb-doped Gadolinium Oxysulfide scintillator screen coupled with a CCD camera. 

The Gamma Profiler is made of two sapphire strip detectors placed perpendicular to each other in order to measure the size of the photon beam in horizontal and transverse plane. 
The $\xi$ can be measured from the photon spot size in the two planes for a linearly polarized beam. 
The Sapphire strip sensors have the dimension of $2\times2$ cm$^2$ with 100~\um of thickness and 100 ~\um of strip pitch. 
These are very radiation hard material and are expected to withstand up to 100 MGy.

\begin{figure}[htbp]
\centering 
\includegraphics[width=0.42\textwidth]{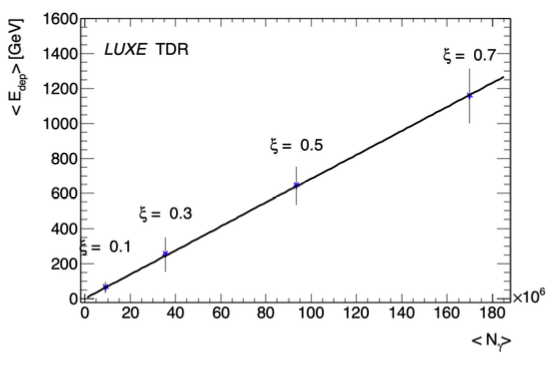}
\caption{Average energy deposition vs the mean number of photons recorded at the backscttering calorimeter. The laser intensity used for the plot is mentioned.}
\label{fig:back}
\end{figure}

The third component of the photon detection system is the Gamma Flux Monitor which consists of 8 lead glass calorimeter blocks around the beam axis with  a radius of $\sim 17$ cm.  
The photons will be back-scattered from the photon dump and it is  found that there is a linear dependence between the energy deposited and the number of incident photons in the backscattering calorimeter.
This linear dependence for different signal sample is shown in~\cref{fig:back}.

\begin{figure}[htbp]
\centering 
\includegraphics[height=1.62in,angle=0]{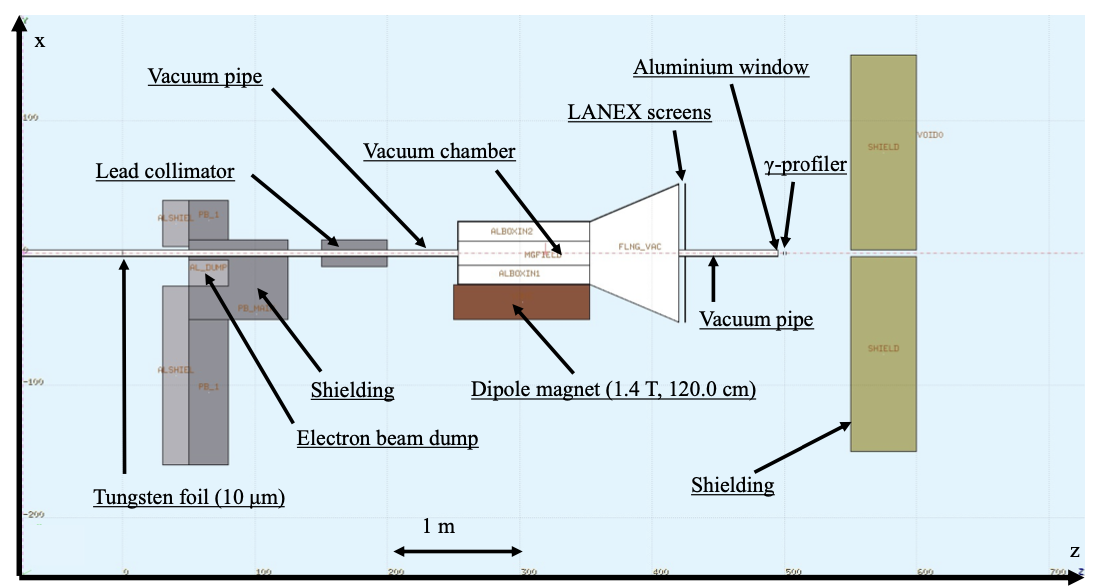}
\includegraphics[height=1.62in,angle=0]{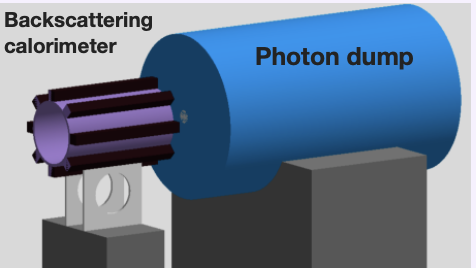}
\caption{The schematic diagram of different components of the  Gamma Spectrometer (left) and backscattering calorimeter with the photon dump (right).}
\label{fig:phds}
\end{figure}

The schematic diagrams of the Gamma Spectrometer and the backscattering calorimeter of the photon detection system are shown in~\cref{fig:phds}. 

\section{Conclusion}
\label{sec:conc}
The LUXE experiment will explore strong field QED regime with multi-terrawatt, high intensity laser and high energetic electron beam from the European XFEL.
This regime is relevant for a variety of astrophysical phenomena, high energy electron accelerators and atomic physics. 
The detectors are designed to achieve a high signal efficiency and a good background rejection for a wide range of signal multiplicity $10^{-2}$ to $10^9$ particles per BX.
Existing technologies for magnets, pixel tracking detectors, Cherenkov counters and calorimeters can be used in the LUXE experiment.
These subsystems will provide with the required efficiency and accuracy in the harsh experimental environment created inside the LUXE experiment.
The installation of this detector is foreseen in 2025 and the data taking period will start from 2026.

\acknowledgments

We thank the DESY technical staff for continuous assistance and the DESY directorate for their strong support and the hospitality they extend to the non-DESY members of the collaboration. This work has benefited from computing services provided by the German National Analysis Facility (NAF) and the Swedish National Infrastructure for Computing (SNIC).


\bibliographystyle{unsrtnat}
\bibliography{jinst-latex-sample}

\end{document}